\documentclass{article}

\usepackage{amsfonts}
\usepackage{amsmath}
\usepackage{amssymb}
\usepackage{amsthm}
\usepackage{breakcites}
\usepackage{caption}
\usepackage[utf8]{inputenc}
\usepackage{longtable}
\usepackage{natbib}
\usepackage{colortbl,xcolor}
\usepackage[a4paper, left=1in, right=1in, bottom=1.2in]{geometry}
\usepackage{graphicx}
\usepackage{hyperref}
\usepackage{subfig}
\usepackage{todonotes}

\usepackage{fancyvrb,newverbs}
\definecolor{cverbbg}{gray}{0.95}

\newenvironment{lcverbatim}
 {\SaveVerbatim{cverb}}
 {\endSaveVerbatim
  \flushleft\fboxrule=1pt\fboxsep=.5em 
  \fcolorbox{black}{cverbbg}{%
    \makebox[\dimexpr\linewidth-2\fboxsep-2\fboxrule][l]{\BUseVerbatim{cverb}}
  }
  \endflushleft
}

\newverbcommand{\cverb}
  {\setbox\verbbox\hbox\bgroup}
  {\egroup\colorbox{cverbbg}{\box\verbbox}}

\newcommand{\tca}{t_{\textrm{ca}}}
\newcommand{\pc}{p_c(t)}
\newcommand{\pvperp}{P_{V^\perp}}
\newcommand{\km}{\textrm{km}}

\DeclareMathOperator{\var}{Var}
\DeclareMathOperator{\expval}{E}

\hypersetup{
  colorlinks,
  linkcolor={red!70!black},
  citecolor={blue!70!black},
  urlcolor={blue!80!black}
}

\begin{document}

\title{Computing the Instantaneous Collision Probability between Satellites using Characteristic Function Inversion}

\author{
  Jason\ Bernstein $|$ bernstein8@llnl.gov \\
  Lawrence Livermore National Laboratory
}

\date{\today}

\setlength{\parskip}{0.5pc}

\maketitle

\begin{center}
{  \setlength{\baselineskip}{.5\baselineskip}
LLNL-TR-863984
\vskip3.0true in
}
\end{center}

\normalsize

\setlength{\parskip}{0.0pc}

\indent

\vspace{-3in}

\begin{abstract}
The probability that two satellites overlap in space at a specified instant of time is called their instantaneous collision probability.
Assuming Gaussian uncertainties and spherical satellites, this probability is the integral of a Gaussian distribution over a sphere.
This paper shows how to compute the probability using an established numerical procedure called characteristic function inversion.
The collision probability in the short-term encounter scenario is also evaluated with this approach, where the instant at which the probability is computed is the time of closest approach between the objects.
Python and R code is provided to evaluate the probability in practice.
Overall, the approach has been established for over fifty years, is implemented in existing software, does not rely on analytical approximations, and can be used to evaluate two and three dimensional collision probabilities.
\end{abstract}

{\renewcommand\arraystretch{1.2}
\noindent\begin{longtable*}{@{}l @{\quad:\quad} l@{}}
\multicolumn{2}{@{}l}{\textbf{Notation}} \\
$q$                         & Distance below which a collision occurs \\
$t$                         & Time at which to compute the collision probability \\
$R=R(t)$                    & Relative position of objects at time $t$ \\
$\mu=\mu(t)$                & Mean of relative position at time $t$ \\
$\Sigma=\Sigma(t)$          & Covariance matrix of relative position at time $t$ \\
$N(\mu,\Sigma)$             & Multivariate normal distribution of $R$ at time $t$ \\
$R'R$                       & Squared distance between objects, or squared separation distance \\
$\pc=p(R'R<q^2)$            & Instantaneous collision probability at time $t$ \\
$\lambda=(\lambda_1,\lambda_2,\lambda_3)'$ & Eigenvalues of $\Sigma$ \\
$A$                         & Eigenvectors of $\Sigma$ \\
\multicolumn{2}{@{}l}{\textbf{Subscripts}} \\
$i=1,2,3$                   & Index for coordinates \\
\multicolumn{2}{@{}l}{\textbf{Acronyms}} \\
ICP                         & Instantaneous collision probability \\
CFI                         & Characteristic function inversion \\
STES                        & Short-term encounter scenario \\
CW                          & Clohessy–Wiltshire
\end{longtable*}}

\section{Introduction}
\label{sec:introduction}

Predicting collisions between space objects, such as satellites or debris, is a critical part of space domain awareness.
Much of the work on this topic computes the probability of a collision in a window of time;
review papers include \cite{alfano2007}, \citet[ch. 8]{klinkrad2006}, \cite{chan2018}, and \cite{li2022}.
Collision scenarios in the literature are typically described as short-term or long-term encounters.
In the short-term encounter scenario (STES), linear, Gaussian assumptions are made that imply the collision probability is the integral of a bivariate Gaussian distribution over a circle \citep{foster1992,akella2000,serra2016}.
In contrast, in a long-term encounter where the relative motion is not approximated as linear, Monte Carlo or other methods are typically employed to compute the collision probability \citep{alfano2009,chan2003,dolado2011}.
In both scenarios, a collision occurs if the distance between the objects, referred to here as the separation distance, falls below a specified threshold in the given time window.

The collision probability when the time window collapses to a single instant is called the instantaneous collision probability (ICP).
Less literature exists for this case compared to the short-term and long-term encounter scenarios.
If the objects are spherical and their relative position is Gaussian distributed, then this probability is the integral of a trivariate Gaussian distribution over a sphere whose radius is the sum of their radii \citep{chan2008}.
Assuming Gaussian distributions, \cite{chan2008} computed the ICP with an equivalent area and an equivalent volume technique.
\cite{zhang2020} computed the ICP using a Gaussian mixture model combined with an equivalent volume approximation.
\cite{masson2023} derived an infinite series for the ICP based on the Laplace transform of the squared separation distance.
Without the Gaussianity assumption, \cite{jones2013} computed the ICP with a surrogate model and Monte Carlo.
In general, methods based on a Gaussian assumption are computationally cheaper but less accurate than methods based on Monte Carlo.

This paper shows how to compute the instantaneous collision probability using characteristic function inversion (CFI), specifically as developed in \cite{davies1973} and \cite{davies1980} for evaluating the distribution of the squared separation distance, which is the generalized chi-square distribution.
The basic idea is that the ICP can be computed by numerically integrating an inversion formula involving the characteristic function of the squared separation distance.
Previously, \cite{bernstein2021} evaluated the conjunction probability in the short-term encounter scenario with this technique.
Hence, some of the material in \cite{bernstein2021} is repeated here, though there are differences in notation.
The collision probability in the STES is also computed with characteristic function inversion in this paper as an example.

In contrast to \cite{zhang2020}, no analytical approximations are needed to use CFI, and so the only source of error is numerical if the object state uncertainties are Gaussian or described by a Gaussian mixture model (GMM).
A GMM extension of the collision model is given in an appendix for completeness, but a single Gaussian model is assumed in the main text for simplicity.
The approach in \cite{masson2023} is similar to ours in that it is based on an inversion of the moment generating function (MGF) of the squared separation distance, and the MGF can be obtained from the characteristic function.
However, the inversion procedure in \cite{masson2023} is significantly different from the method developed in the papers by Davies.
There is also literature on this probability outside the space object context, where it is called an offset hitting probability.
For example, \cite{grubbs1964} approximated the probability using normal and chi-square distributions and \cite{cacoullos1984} noted the distribution of the squared separation distance is the generalized chi-square distribution.

The formula for the instantaneous collision probability also applies to conjunction and other close proximity events.
That is, the formula for computing the instantaneous collision probability is the same as the formula for the probability that two objects are any distance apart, not just a threshold distance that implies a collision occurs.
This paper uses the term collision instead of a more general term to agree with previous work on this problem and for simplicity.

The outline of this paper is as follows.
Section \ref{sec:model} describes the instantaneous collision model and the corresponding probability to be computed.
Section \ref{sec:gcsd} derives the generalized chi-square distribution of the squared separation distance.
Section \ref{sec:cfi} describes the characteristic function inversion procedure for computing the collision probability.
Section \ref{sec:code} gives \verb+R+ and \verb+Python+ code for computing the probability.
Section \ref{sec:examples} provides examples, including the short-term encounter scenario and the Clohessy-Wiltshire model.
Section \ref{sec:bound} gives an upper bound on the probability that is slightly faster to compute than the actual probability.
Section \ref{sec:discussion} provides a discussion of the assumptions and interpretation of the probability.
Section \ref{sec:conclusion} concludes the paper and suggests possible future work.

\section{Collision Model and Probability}
\label{sec:model}

This section presents the assumed collision model and gives an expression for the instantaneous collision probability.
The relative position of the objects at time $t$ is denoted $R=(R_1,R_2,R_3)'$ and assumed to be normally distributed with mean $\mu=(\mu_1,\mu_2,\mu_3)'$ and positive definite covariance matrix $\Sigma$, denoted
\begin{equation}
\label{distnR}
R\sim N(\mu,\Sigma).
\end{equation}
This model for the relative position is obtained by assuming that object positions are jointly Gaussian distributed.
No assumption on the independence of the objects is made in this model, so the object positions could be correlated.
Note also that the relative velocity of the objects is not needed to compute the ICP since time is fixed in this situation.

Assuming spherical objects, a collision occurs if the distance between the objects is less than the sum of their radii;
equivalently, the objects collide if the squared separation distance, $R'R$, is less than the square of the sum of their radii.
Hence, if $q$ is the sum of the radii of the objects, then a collision occurs if $R'R<q^2$.
The instantaneous collision probability is then
\begin{align}
\label{ICP}
\pc&=p(R'R<q^2)\\
&=\int_{\{r|r'r<q^2\}}\exp\left(-\frac{1}{2}(r-\mu)'\Sigma^{-1}(r-\mu)\right)\,dr,
\end{align}
where $R'R=R_1^2+R_2^2+R_3^2$ is the squared separation distance and $r=(r_1,r_2,r_3)'$.
The goal is to compute the instantaneous collision probability, $\pc$, by evaluating this integral.
Note that $R$, $\mu$, and $\Sigma$ depend on the time, $t$, but $t$ is supressed in the notation when possible.

The variable $q$ can represent a collision distance threshold, a close-approach threshold, or some other distance of interest, and this interpretation does not change the calculation of the probability in \eqref{ICP}.
Here, $q$ is called a collision distance threshold since that is how it is usually interpreted in the literature.
However, the probability $\pc$ is not necessarily the collision probability at time $t$ as discussed in section~\ref{sec:discussion}.

The collision model can also be generalized by assuming a Gaussian mixture distribution for the relative position, as described in \cite{zhang2020} and the references therein.
In this case, the ICP is a weighted sum of terms of the form in \eqref{ICP}, where the weights are from the mixture model.
We assume a single Gaussian model to avoid complexity, but give the mixture model details in the appendix for completeness.

\section{Generalized Chi-Square Distribution}
\label{sec:gcsd}

This section shows that the squared separation distance has the generalized chi-square distribution, which is the distribution of a linear combination of independent, non-central chi-square random variables.
The work is thus to compute the coefficients and non-centrality parameters of these chi-square random variables.
These quantities are obtained from $\mu$ and the eigendecomposition of the combined covariance matrix, $\Sigma$.
The next section then shows how to evaluate this distribution to compute the instantaneous collision probability.
References on decompositions of quadratic forms of Gaussian random vectors include \cite{duchesne2010} and \citet[ch. 5]{rencher2008}.

Let $A$ denote the matrix whose columns are the eigenvectors of $\Sigma$ and let $\lambda=(\lambda_1,\lambda_2,\lambda_3)'$ denote the eigenvalues of $\Sigma$, so that
\begin{equation}
\Sigma=ADA',
\end{equation}
where $D$ is a diagonal matrix whose diagonal elements are the $\lambda_i$.
The eigenvectors point in the orthogonal directions of maximal variance and the $\lambda_i$ are those variances.
The distribution of the relative position can therefore be written as
\begin{equation}
R\sim N(\mu, ADA').
\end{equation}
The rotated and scaled relative position vector,
\begin{equation}
Y=D^{-1/2}A'R,
\end{equation}
has unit variances and correlations equal to zero, so that
\begin{equation}
Y\sim N(\nu, I_3),
\end{equation}
where $I_3$ is the $3\times3$ identity matrix and the mean of $Y$ is
\begin{equation}
\nu=D^{-1/2}A'\mu.
\end{equation}
The squared separation distance can then be written as
\begin{align}
\label{quad_form}
R'R&=Y'DY\\
\label{lincomb}
&=\sum_{i=1}^3\lambda_iY_i^2.
\end{align}
Note that $Y_i^2$ has a chi-square distribution with one degree of freedom and non-centrality parameter $\delta_i=\nu_i^2$, denoted
\begin{equation}
Y_i^2\sim \chi^2_1(\delta_i),
\end{equation}
for $i=1,2,3$.
Furthermore, the $Y_i$ are independent since their covariances are zero, which implies that the $Y_i^2$ are independent.
Thus, the squared separation distance is a linear combination of independent and non-central chi-square random variables.

The squared separation distance in \eqref{quad_form} is said to have the generalized chi-square distribution.
When all the $\lambda_i=1$, the distribution reduces to the non-central chi-square distribution with three degrees of freedom and non-centrality parameter $\delta_1 + \delta_2 + \delta_3$.
More generally if $\lambda_i=c$ for some positive constant $c$, then the distribution of the squared separation distance can be evaluated with a non-central chi-square distribution function.
There are also intermediate cases where a subset of the $\lambda_i$ are equal, but we do not discuss those cases in detail here.

The statistical properties of the generalized chi-square distribution are well-known and could be used for conjunction screening.
In particular, the mean and variance of the squared separation distance are given by equations 5.4 and 5.9 of \citet{rencher2008} as
\begin{align}
\expval[R'R]&=\textrm{tr}(\Sigma)+\mu'\mu\\
\var[R'R]&=2\textrm{tr}(\Sigma\Sigma) + 4\mu'\Sigma\mu,
\end{align}
where $\textrm{tr}(\cdot)$ is the matrix trace function.
Equivalently, from \eqref{lincomb}, the mean and variance of the squared separation distance are
\begin{align}
\expval[R'R]&=\sum_{i=1}^3\lambda_i(1+\delta_i)\\
\var[R'R]&=2\sum_{i=1}^3\lambda_i^2(1+2\delta_i).
\end{align}
The mean of $R'R$ can quickly be computed and compared to $q^2$ to informally assess the potential for a collision.
Actual collision probabilites can be computed using characteristic function inversion, as described next.

\section{Characteristic Function Inversion}
\label{sec:cfi}

This section shows how to compute the ICP by inverting the characteristic function of the squared separation distance.
The inversion algorithm and its application to the generalized chi-square distribution are from \cite{davies1973,davies1980}.
Only the characteristic function and the inversion formula for computing the ICP from the characteristic function are given here.
Additional details can be found in the papers by Davies and in the literature on characteristic function inversion.

Following \citet[sec. 4]{davies1973}, the characteristic function of the squared separation distance in \eqref{lincomb} is
\begin{equation}
\label{chf}
\phi(s)=\prod_{i=1}^3\phi_i(\lambda_is),
\end{equation}
where $\iota=\sqrt{-1}$, $s\in\mathbb{R}$, and
\begin{equation}
\phi_i(s)=(1-2\iota s)^{-1/2}\exp\left(\frac{\iota\delta_is}{1-2\iota s}\right)
\end{equation}
is the characteristic function of $Y_i^2$.
That is, $\phi_i(s)$ is the characteristic function of a non-central chi-square random variable with one degree of freedom and non-centrality parameter $\delta_i$.
Further note that $\phi(s)$ is the product of the characteristic functions of the $Y_i^2$ since the $Y_i$ are independent.
The characteristic function of the squared separation distance can also be written in terms of $\mu$ and $\Sigma$ following \citet[thrm. 5.2b]{rencher2008}.
However, software for inverting the characteristic function takes $\lambda$ and $\delta$ as inputs instead of $\mu$ and $\Sigma$, so the former parameterization is used here.

Following \cite{davies1973}, the instantaneous collision probability can be computed using the characteristic function inversion formula in \cite{gil1951} as
\begin{equation}
\label{cfi}
\pc=\frac{1}{2}-\int_{-\infty}^{\infty}\textrm{Im}\left(\frac{\phi(s)\exp(-\iota sq^2)}{2\pi s}\right)\,ds.
\end{equation}
To gain intuition into this equation, note that the probability that the two objects have the exact same position is zero, so that $\pc=0$ and the integral equals 1/2 when $q=0$.
Similarly, a collision is more likely to occur as the radius of the spherical conjunction domain increases, so that $\pc\rightarrow1$ and the integral converges to -1/2 as $q\rightarrow\infty$.
Since $\pc$ increases with $q$, the integral decreases with $q$ from 1/2 at $q=0$ to -1/2 as $q\rightarrow\infty$.

There are two key points about the integral representation of the instantaneous collision probability in \eqref{cfi}.
First, the integral is univariate, whereas the original expression for $\pc$ in \eqref{ICP} is a trivariate integral.
Second, \cite{davies1973} and \cite{davies1980} give error bounds on evaluating \eqref{cfi} numerically with the trapezoidal rule.
Hence, the instantaneous collision probability can be computed as a one-dimensional integral with the trapezoidal rule up to a specified error tolerance.
Furthermore, the method has been established for evaluating the generalized chi-square distribution for over fifty years and is available in open-source \verb+R+ and \verb+Python+ packages.
The software is described in the following section and then is applied to sample scenarios.

\section{Software Implementation}
\label{sec:code}

This section provides \verb+R+ and \verb+Python+ functions for computing the ICP using the CFI algorithm of \citet{davies1980}.
From sections \ref{sec:gcsd} and \ref{sec:cfi}, the process is to compute the $\lambda$ and $\delta$ parameters from $\mu$ and $\Sigma$, and then to compute the ICP from \eqref{cfi}.
An \verb+R+ function for estimating the ICP using Monte Carlo is also provided to check the probabilities are correct.
A \verb+Matlab+ software suite for evaluating the generalized chi-square distribution is described in \cite{das2024}.

\subsection{R Code}
\label{ssec:R}

This \verb+R+ code loads two required packages, \verb+CompQuadForm+ \citep{duchesne2010} and \verb+mvnfast+ \citep{fasiolo2014}.
The \verb+CompQuadForm+ package is used to compute the instantaneous collision probability with characteristic function inversion.
The \verb+mvnfast+ package is used to sample multivariate normal random variables for computing the ICP using Monte Carlo.

\begin{lcverbatim}
# packages needed for analysis
library(CompQuadForm)
library(mvnfast)
\end{lcverbatim}

This \verb+R+ function computes the ICP using the \verb+davies+ function in the \verb+CompQuadForm+ package.
In addition to $q$, $\mu$, and $\Sigma$, the \verb+davies+ function has two arguments, \verb+lim+ and \verb+acc+, that give an upper bound on the number of terms in the numerical integration and the error tolerance, respectively.
The function documentation suggests \verb+lim+ values between 1,000 and 50,000 and \verb+acc+ values between 0.00005 and 0.001.
The \verb+stes+ argument is a Boolean flag for computing the collision probability in the short-term encounter scenario.
When \verb+stes+ is set to True, the arguments mu and sigma are the mean and covariance matrix of the relative position at the time of closest approach.
A detailed description of the scenario and how these arguments are computed is given in section \ref{ssec:stes}.

\begin{lcverbatim}
comp_icp_cfi <- function(q, mu, sigma, stes, lim = 5e4, acc = 1e-6) {
  # compute instantaneous Pc using characteristic function inversion
  eig <- eigen(sigma)
  A <- eig$vectors
  lambda <- eig$values
  if (stes) {
    A <- A[, 1:2]
    lambda <- lambda[1:2]
  }
  nu <- (t(A) 
  delta <- nu**2
  pc <- 1 - CompQuadForm::davies(q**2,
    lambda = lambda, delta = delta,
    lim = lim, acc = acc
  )$Q
  return(pc)
}
\end{lcverbatim}

This \verb+R+ function estimates the instantaneous collision probability using Monte Carlo.

\begin{lcverbatim}
comp_icp_mc <- function(q, mu, sigma, n = 1e7, seed = 1) {
  # compute instantaneous Pc using Monte Carlo
  set.seed(seed)
  R <- mvnfast::rmvn(n, mu, sigma)
  sq_miss_dist <- rowSums(R * R)
  pc <- mean(sq_miss_dist < q**2)
  return(pc)
}
\end{lcverbatim}

\subsection{Python Code}
\label{ssec:python}

In \verb+Python+, the \verb+chi2comb+ \citep{chi2comb} and \verb+numpy+ packages are used to compute the ICP.

\begin{lcverbatim}
# packages needed for analysis
from chi2comb import ChiSquared as chisq, chi2comb_cdf
import numpy as np
\end{lcverbatim}

After loading these packages, this function can be used to compute the instantaneous collision probability.

\begin{lcverbatim}
def comp_icp_cfi(q, mu, sigma, stes, lim=50000, atol=1e-6):
  # compute instantaneous Pc using characteristic function inversion
  Lambdas, A = np.linalg.eig(sigma)
  if stes:
    Lambdas = Lambdas[:2]
    A = A[:, :2]
  nu = A.T @ mu / np.sqrt(Lambdas)
  deltas = nu**2
  Y = [chisq(Lambda, ncent=delta, dof=1) for Lambda, delta in zip(Lambdas, deltas)]
  pc = chi2comb_cdf(q**2, Y, gcoef=0, lim=lim, atol=atol)[0]
  return pc
\end{lcverbatim}

The \verb+gcoef+ argument in the \verb+chisq+ function is the coefficient of a Gaussian random variable and is set to zero since there is no such term for this problem.

\section{Examples}
\label{sec:examples}

This section gives examples of computing instantaneous collision probabilities using characteristic function inversion.
Section \ref{ssec:synthetic} gives synthetic scenarios to illustrate the method and provides values for benchmarking.
Section \ref{ssec:stes} computes the collision probability in the short-term encounter scenario.
Last, section \ref{ssec:cw} considers an example with the Clohessy–Wiltshire (CW) model.

\subsection{Synthetic Scenarios}
\label{ssec:synthetic}

\begin{figure}[h]
\centering
\includegraphics[scale=0.5]{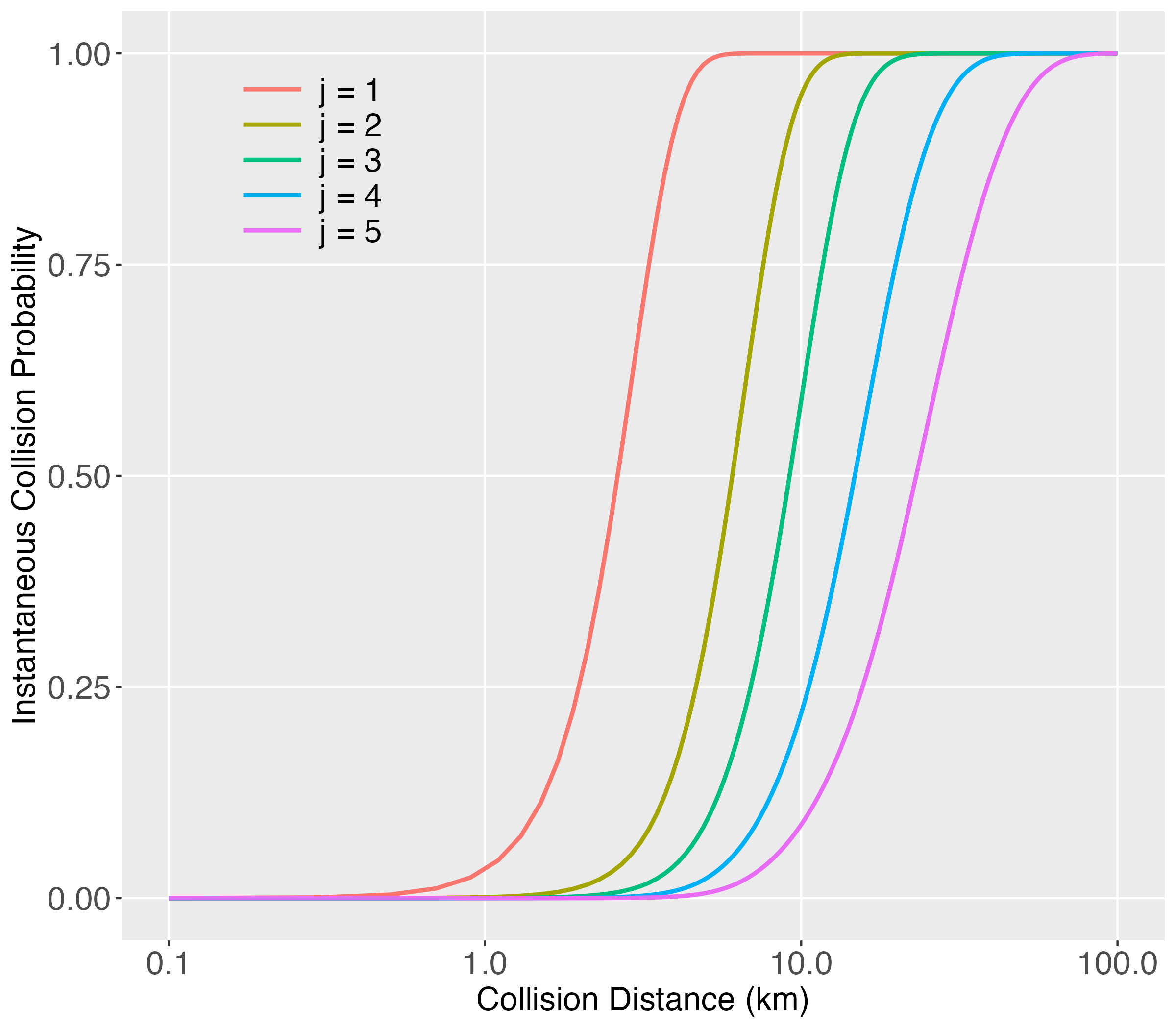}
\caption{Instantaneous collision probability versus the distance below which a collision occurs, $q$, for the relative mean vectors and covariance matrices from \eqref{synthparams}.}
\label{fig:pc}
\end{figure}

Instantaneous collision probabilities are computed for five synthetic scenarios as a function of the collision distance.
For these five scenarios, the $j$th mean vector and covariance matrix are
\begin{equation}
\label{synthparams}
\mu_j=
\begin{bmatrix}
j\\
2j\\
j+(-1)^j
\end{bmatrix}
\qquad\textrm{and}\qquad
\Sigma_j=\frac{j}{2}\begin{bmatrix}
1    & 0.5  & 0.25 \\
0.5  & 2    & -0.7 \\
0.25 & -0.7 & 3
\end{bmatrix}^j
\end{equation}
for $j=1,2,3,4,5$.
The instantaneous collision probabilities as a function of $q$ for these cases are shown in Figure \ref{fig:pc}.
In general, the collision probability decreases in $j$ for fixed $q$ since the distance between the objects increases with $j$.
A subset of the instantaneous collision probabilities, computed with CFI and estimated with Monte Carlo, are shown in Table 1.
These values are provided for benchmarking alternative methods again CFI and Monte Carlo.
The Monte Carlo estimates are based on a sample size of one million.

\begin{table}[ht]
\centering
\begin{tabular}{|c|c|c|c|}
  \hline
$j$ & $q$ & $p_c(t)$ (CFI) & $p_c(t)$ (Monte Carlo) \\ 
  \hline
 1 &    3 & 0.647 & 0.648 \\ 
   \rowcolor[gray]{0.95} 1 &    4 & 0.913 & 0.913 \\ 
   1 &    5 & 0.989 & 0.989 \\ 
   \rowcolor[gray]{0.95} 2 &    3 & 0.043 & 0.043 \\ 
   2 &    4 & 0.120 & 0.120 \\ 
   \rowcolor[gray]{0.95} 2 &    5 & 0.256 & 0.256 \\ 
   3 &    3 & 0.025 & 0.025 \\ 
   \rowcolor[gray]{0.95} 3 &    4 & 0.053 & 0.053 \\ 
   3 &    5 & 0.096 & 0.096 \\ 
   \rowcolor[gray]{0.95} 4 &    3 & 0.008 & 0.008 \\ 
   4 &    4 & 0.016 & 0.016 \\ 
   \rowcolor[gray]{0.95} 4 &    5 & 0.028 & 0.028 \\ 
   5 &    3 & 0.005 & 0.005 \\ 
   \rowcolor[gray]{0.95} 5 &    4 & 0.010 & 0.010 \\ 
   5 &    5 & 0.017 & 0.017 \\ 
   \hline
\end{tabular}
\caption{Instantaneous collision probabilities computed using characteristic function inversion and Monte Carlo for the synthetic scenarios.} 
\label{tab:pc}
\end{table}

\subsection{Short-term Encounter Scenario}
\label{ssec:stes}

The collision probability in the short-term encounter scenario (STES) is now evaluated as an instantaneous collision probability.
Several references for the short-term encounter scenario are given in section~\ref{sec:introduction}.
The assumptions are that the relative motion of the objects is linear, the relative velocity is known and constant, and the initial relative position is normally distributed.
The key insight is that the collision probability in the STES is the ICP at the time of closest approach between the objects.
Hence, the main task is to compute the mean and covariance matrix of the relative position at the time of closest approach under the STES assumptions, from which the ICP can be evaluated using characteristic function inversion.

Let $R$ denote the initial relative position with $R\sim N(\mu,\Sigma)$ and let $V$ denote the initial relative velocity.
The relative position of the objects at time $t$ is $R+tV$, which is minimized at the time of closest approach,
\begin{equation}
\label{tca}
\tca=-R'V/(V'V).
\end{equation}
The miss vector, or relative position at the time of closest approach, is
\begin{align}
R+\tca V&=\left(I_3-\frac{VV'}{V'V}\right)R \\
&=\pvperp R,
\end{align}
where $\pvperp=I_3-(VV')/(V'V)$ is a projection matrix since it projects $R$ onto the plane orthogonal to $V$.
This plane is typically called the encounter or conjunction plane.
By linearity, the miss vector is normally distributed,
\begin{equation}
\label{Rtca}
\pvperp R\sim N(\pvperp\mu,\pvperp\Sigma\pvperp),
\end{equation}
so the collision probability in the STES is
\begin{equation}
p_c(\tca)=p(R'\pvperp R<q^2)
\end{equation}
since $\pvperp'\pvperp=\pvperp$.
The collision probability can then be computed using CFI with the mean and covariance matrix in \eqref{Rtca}.
\cite{bernstein2021} gives additional details and references for this derivation.

There are three points to emphasize about the instantaneous collision probability in the STES:
\begin{enumerate}
\item The time of closest approach in the STES, $\tca$, is random since $R$ is Gaussian distributed.
In the initial discussion of the instantaneous collision probability, the time, $t$, was treated as fixed.
\item The covariance matrix of the miss vector in \eqref{Rtca} has rank 2 since $\pvperp$ has rank 2, implying that $\lambda_3=0$.
Previously, the covariance matrix of the relative position, $\Sigma$, was assumed to have rank 3.
The \verb+R+ and \verb+Python+ functions are therefore modified in the STES by discarding $\lambda_3$ and the third column of $A$.
The matrix $D$ is then a diagonal matrix with diagonal $(\lambda_1,\lambda_2)'$ and $\nu$ is a two-dimensional vector.
\item As noted in \eqref{ICP}, the ICP is the integral of a trivariate Gaussian probability density function (PDF) over a sphere.
However, at the time of closest approach, the ICP is the integral of a bivariate Gaussian PDF over a circle since $\lambda_3=0$.
The circle is contained in the conjunction plane defined above.
\end{enumerate}
Overall, the ICP at an arbitrary time or at the time of closest approach in the STES reduces to a one-dimensional integral that can be evaluated with CFI.
An advantage of using CFI for evaluating collision probabilities is thus that only one method is needed for the two separate situations.

A numerical example helps illustrate the computations.
Let the initial mean and covariance matrix of the relative position be
\begin{equation}
\mu=\begin{pmatrix}
5\\
10\\
15
\end{pmatrix},\qquad
\Sigma=\begin{pmatrix}
9 & 37 & 18\\
37 & 165 & 68\\
18 & 68 & 86
\end{pmatrix},
\end{equation}
and let $V=(-2, 0, 3)'$ and $q=5$.
The mean and covariance matrix of the relative position at the time of closest approach are
\begin{equation}
\pvperp\mu=\begin{pmatrix}
10.38\\
10.00\\
6.92
\end{pmatrix},\qquad
\pvperp\Sigma\pvperp=\begin{pmatrix}
34.14 & 57.00  & 22.76\\
57.00 & 165.00 & 38.00\\
22.76 & 38.00  & 15.17
\end{pmatrix}.
\end{equation}
Note that $\lambda=(196.8,17.49,0)'$, where $\lambda_3=0$ since the covariance matrix is provided at the time of closest approach.
The collision probability at the time of closest approach is computed with the code in section~\ref{sec:code} as 0.038.

\subsection{Clohessy–Wiltshire Model}
\label{ssec:cw}

The Clohessy–Wiltshire (CW) model describes the relative motion of two objects in a rendezvous \citep{clohessy1960}.
Under certain assumptions on the satellite orbits, the relative motion is linear in the initial relative position.
Assuming that the initial relative position is Gaussian distributed then implies that the relative position at future times is also Gaussian distributed, and therefore the instantaneous collision probability can be evaluated with characteristic function inversion.
The CW model is thus a useful example for evaluating collision or close-approach probabilities.
Other possible sources of uncertainty, such as process noise or uncertainty in the relative velocity, are ignored.

Following \citet[sec. 7.4]{curtis2013}, the relative position of the objects in the CW model at time $t$ is
\begin{equation}
\label{cwR}
R(t)=\Phi_{rr}(t)R+\Phi_{rv}(t)V,
\end{equation}
where $R$ and $V$ are the initial relative position and relative velocity, respectively,
\begin{equation}
\Phi_{rr}(t)=
\begin{pmatrix}
4-3\cos(nt) & 0 & 0 \\
6(\sin(nt)-nt) & 1 & 0 \\
0 & 0 & \cos(nt)
\end{pmatrix},
\end{equation}
and
\begin{equation}
\Phi_{rv}(t)=\frac{1}{n}
\begin{pmatrix}
\sin(nt) & 2(1-\cos(nt)) & 0 \\
2(\cos(nt)-1) & 4\sin(nt)-3nt & 0 \\
0 & 0 & \sin(nt)
\end{pmatrix}.
\end{equation}
The initial relative position is assumed to satify $R\sim N(\mu,\Sigma)$, so that
\begin{equation}
\label{cwdist}
R(t)\sim N(\mu_{cw}(t),\Sigma_{cw}(t)),
\end{equation}
where from \eqref{cwR},
\begin{equation}
\mu_{cw}(t)=\Phi_{rr}(t)\mu+\Phi_{rv}(t)V
\end{equation}
and
\begin{equation}
\Sigma_{cw}(t)=\Phi_{rr}(t)\Sigma\Phi_{rr}'
\end{equation}
are the mean and covariance matrix of the relative position at time $t$.
The instantaneous collision probability at time $t$ can then be computed with CFI from the distribution in \eqref{cwdist}.

\begin{figure}[h]
\centering
\includegraphics[scale=0.5]{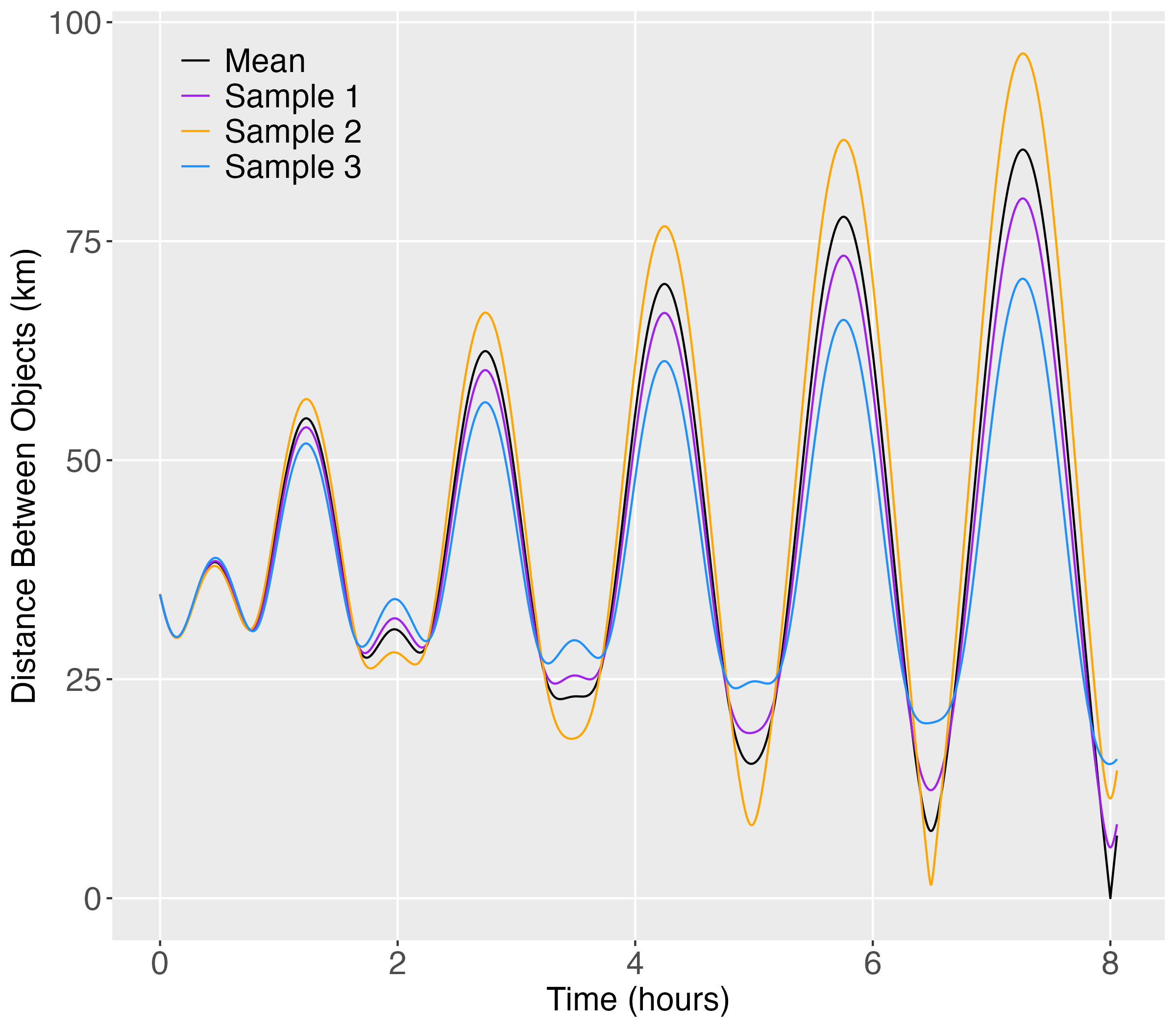}
\caption{Relative position from the CW model over time for the mean initial relative position and for three random initial relative positions.}
\label{fig:cw_samples}
\end{figure}

As an illustration, example 7.4 from \cite{curtis2013} is modified to include uncertainty in the initial relative position.
Specifically, the mean initial relative position and velocity, and the covariance matrix of the initial relative position, are taken to be
\begin{equation}
\mu=
\begin{pmatrix}
20 \\
20 \\
20
\end{pmatrix}\,\km,
\qquad
V=
\begin{pmatrix}
0.00930458 \\
-0.0467472 \\
0.00798343
\end{pmatrix}\,\frac{\km}{\textrm{sec}},
\qquad
\Sigma=
\begin{pmatrix}
0.003 & 0 & 0 \\
0 & 0.003 & 0 \\
0 & 0 & 0.003
\end{pmatrix}\,\km^2.
\end{equation}
If the initial relative position is $\mu$, then the objects reach their point of closest approach eight hours later.
However, if the actual initial relative position is $R\sim N(\mu,\Sigma)$, then the point of closest approach can occur at a different time.
For example, Figure \ref{fig:cw_samples} shows the relative position over time for four different initial relative positions. 
In this plot, the second trajectory attains its point of closest approach around 6.5 hours after the initial time.

\begin{figure}[h]
\centering
\includegraphics[scale=0.5]{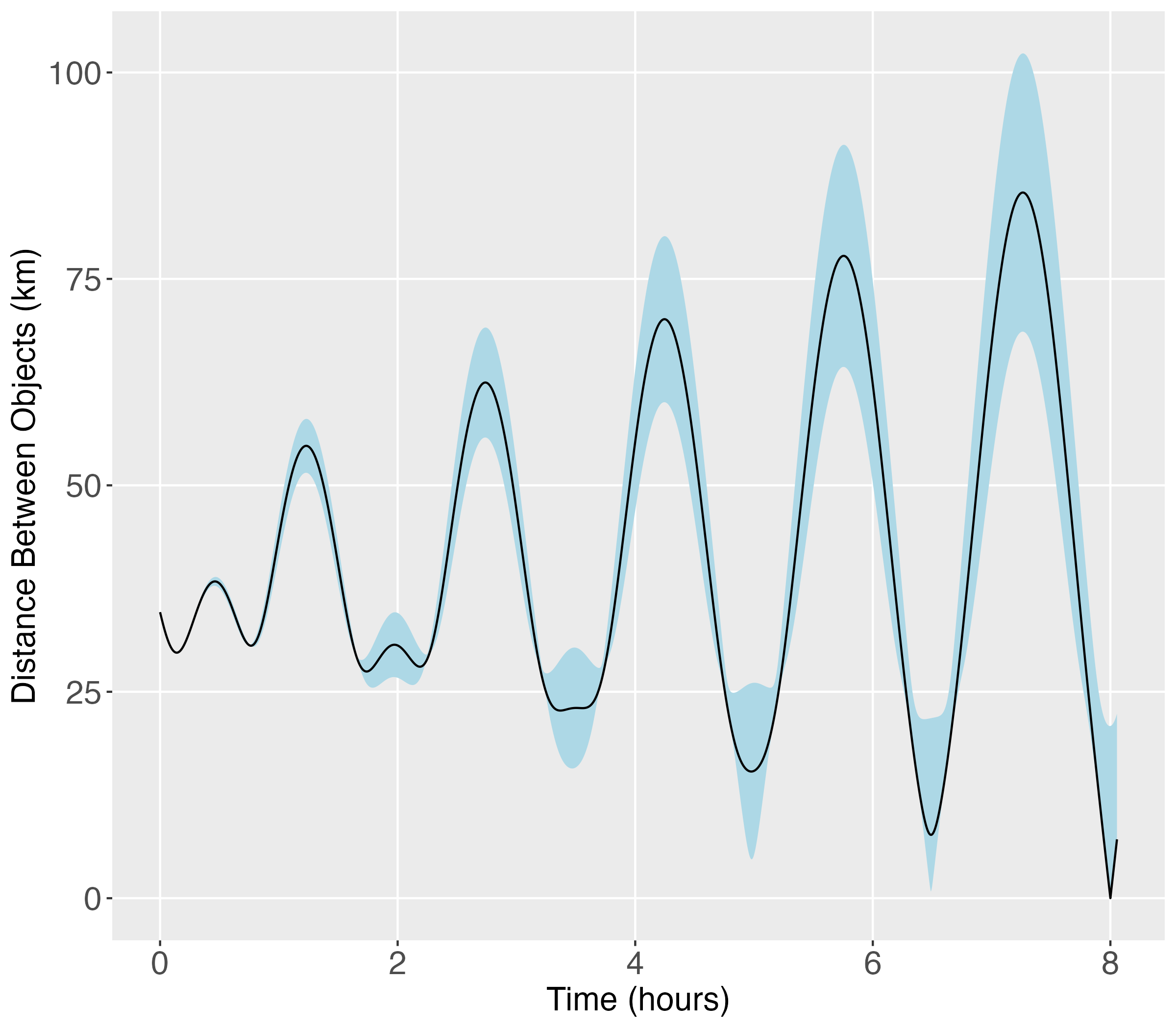}
\caption{Relative position over time starting at the mean initial relative position (black line) and a 90\% probability interval assuming Gaussian uncertainty in the initial relative position (blue shaded area).}
\label{fig:cw_confint}
\end{figure}

Figure \ref{fig:cw_confint} shows the relative position over time starting from the mean initial relative position and the corresponding 90\% probability interval from the Gaussian uncertainty in the initial relative position.
The 5th and 95th percentiles of the generalized chi-square distribution are computed quickly using bisection.
Computing these percentiles with Monte Carlo is more computationally expensive and unnecessary.
The main reason to use Monte Carlo for this problem is to gain intuition or check results obtained with CFI.

\section{Upper Bound on Instantaneous Collision Probability}
\label{sec:bound}

Computing the instantaneous collision probability with CFI can be slow if the accuracy required is high.
However, an upper bound on the probability can be computed quickly by integrating the relative position distribution over a bounding box.
Specifically, the upper bound is the probability that the objects fall within a box centered at the origin and with side lengths equal to $2q$, since this box contains the spherical conjunction domain.
This upper bound on the probability is
\begin{equation}
\label{upperbound}
\prod_{i=1}^3\left(\Phi(q;\lambda_i^{1/2}\nu_i,\lambda_i)-\Phi(-q;\lambda_i^{1/2}\nu_i,\lambda_i)\right),
\end{equation}
where $\Phi(x;\lambda_i^{1/2}\nu_i,\lambda_i)$ is the cumulative distribution function of a normal random variable with mean $\lambda_i^{1/2}\nu_i$, variance $\lambda_i$, and evaluated at $x$.
A similar upper bound for the conjunction probability in the STES is given in section 3.3 of \cite{bernstein2021}.
Note also that $\lambda_3=0$ in the STES as noted in section~\ref{ssec:stes}, in which case the third term in \eqref{upperbound} is equal to one.

\begin{figure}[h]
\centering
\includegraphics[scale=0.5]{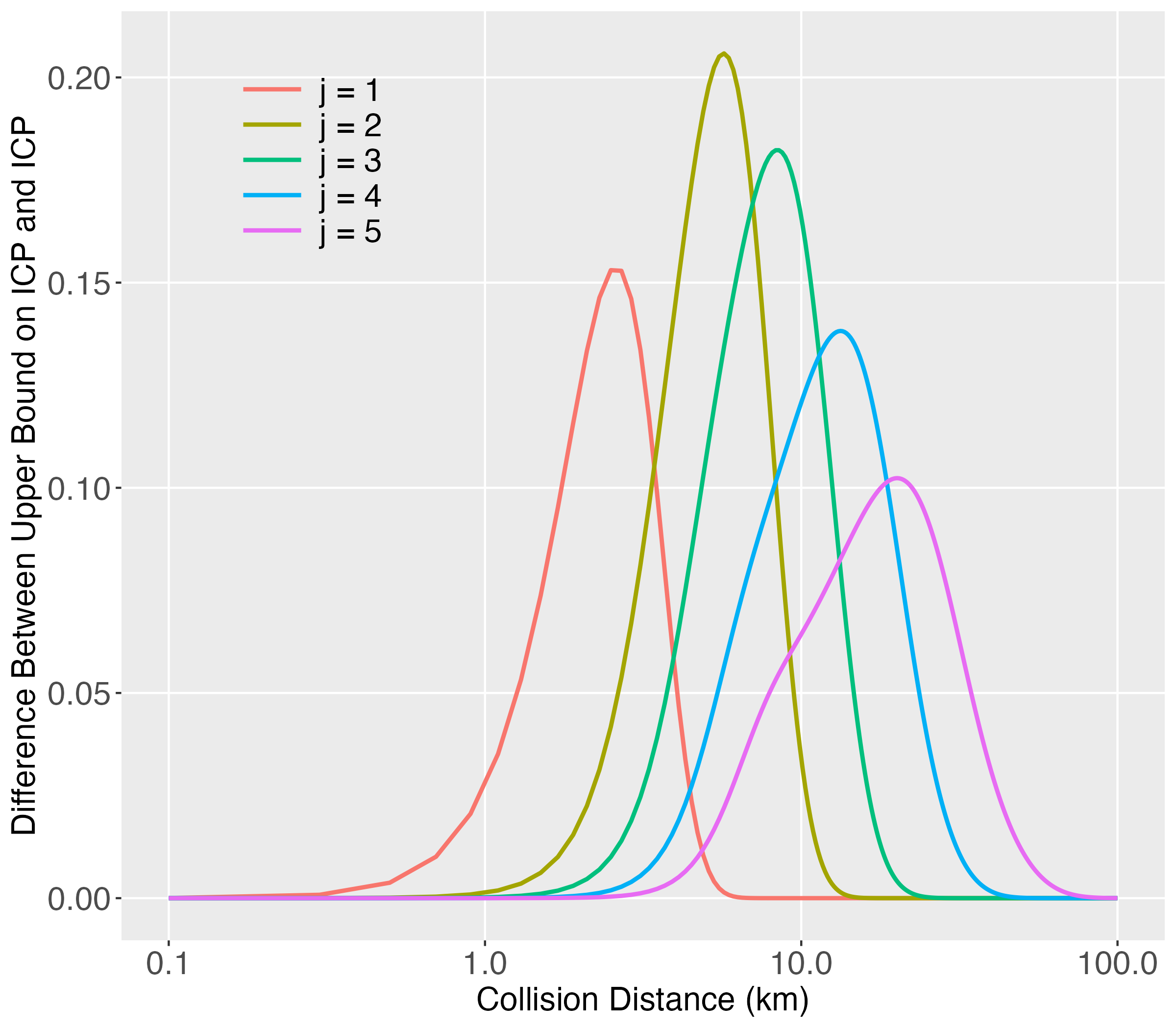}
\caption{Difference between the upper bound on the instantaneous collision probability and the actual instantaneous collision probability. The values of $j$ correspond to those in Figure \ref{fig:pc}.}
\label{fig:pc_upper_bound}
\end{figure}

Figure~\ref{fig:pc_upper_bound} shows the difference between the upper bound in \eqref{upperbound} and the actual instantaneous collision probability. 
The maximum difference exceeds 0.1 for each case in the figure and exceeds 0.2 when $j=2$ and $q$ is near 6.
The upper bound is therefore conservative for these cases when $q$ is approximately between $1\;\km$ and $100\;\km$, suggesting the actual probability should be computed for these values of $q$ instead.

\begin{figure}[h]
\centering
\includegraphics[scale=0.5]{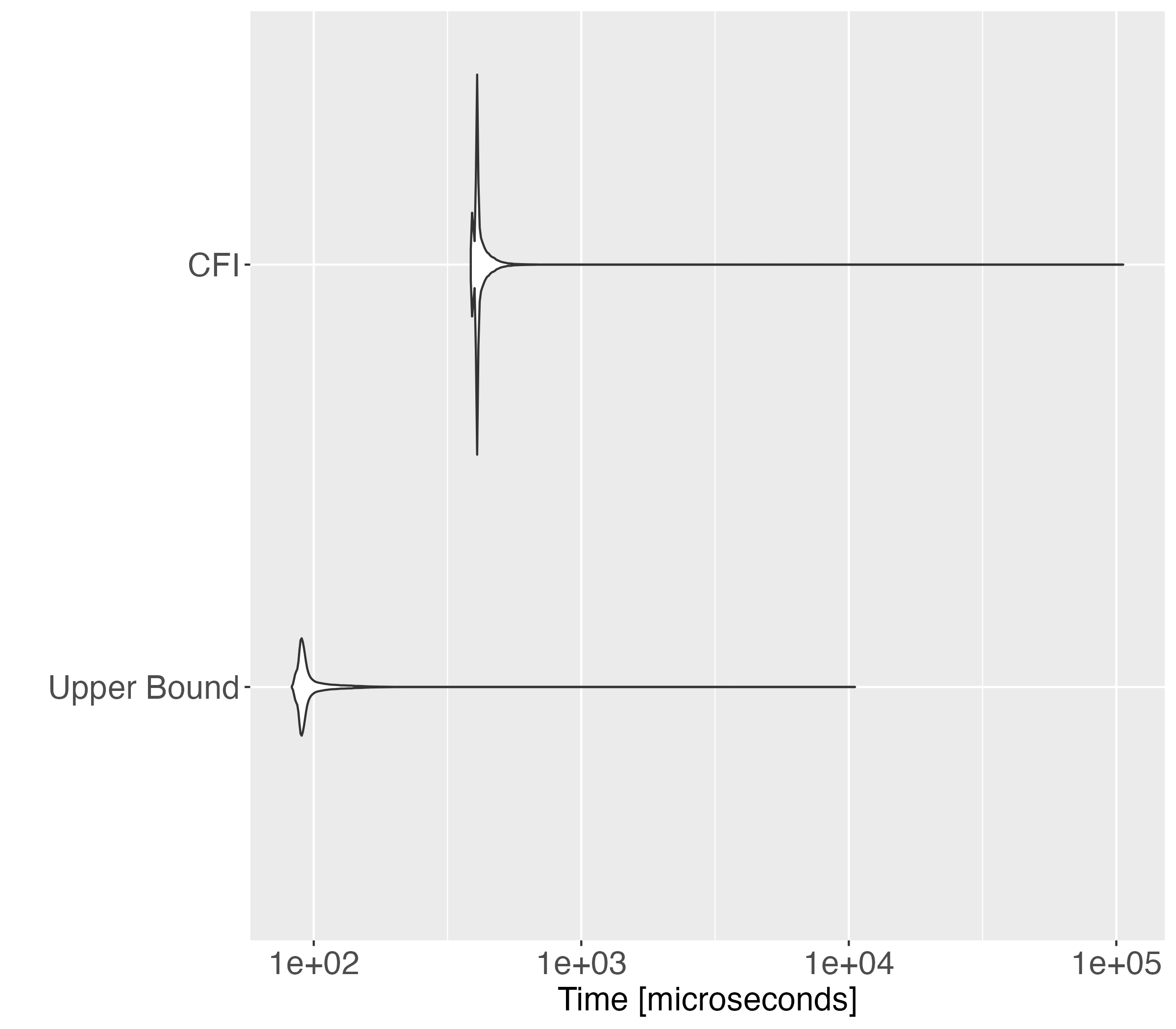}
\caption{Time to compute the instantaneous collision probability using CFI and the upper bound in \eqref{upperbound}. Results are shown for one-hundred thousand evaluations, $q=2$, $\nu=(-1.06,0.55,1.86)'$, $\lambda=(3.52,1.59,0.45)'$, and $\pc=0.196$.}
\label{fig:benchmark}
\end{figure}

Figure \ref{fig:benchmark} compares the time to evaluate the instantaneous collision probability and its upper bound.
For one-hundred thousand evaluations, the average time to evaluate the upper bound is approximately 93 microseconds.
The average time to compute the actual probability with CFI and an accuracy of $10^{-6}$ is approximately 399 microseconds.
Hence, the average time to evaluate the upper bound is about four times less than the average time to evaluate the exact probability.
In this case, the actual probability is 0.196 and the upper bound is 0.351.

Further increasing the accuracy of the characteristic function inversion calculation will generally increase the evaluation time.
Since the upper bound is faster to compute than the actual probability and high accuracy is likely not needed in many cases, a reasonable approach to assessing instantaneous collision risk is to first compute the upper bound and then to compute the more accurate probability using CFI if the upper bound is above some pre-specified threshold.
This approach is efficient if the upper bound is usually not above the threshold that leads to evaluation of the actual probability.
Further investigation into the trade-off in compute time versus accuracy could better inform when to compute the upper bound before computing the actual probability with characteristic function inversion.

\section{Discussion}
\label{sec:discussion}

Care is required in interpreting Figure~\ref{fig:cw_confint} because the instantaneous collision probabilities defined in \eqref{ICP} are plotted over time, but they are not the actual collision probabilities at those times.
The reason is that the ICP in \eqref{ICP} does not condition on the absence of collisions in $(0,t)$, which is a requirement for there to be a collision at time $t$.
For example, there cannot be a collision two, four, and six hours after the initial time.
The actual collision probability at time $t$, given no collisions over $(0,t)$, is therefore $p(R(t)'R(t)<q^2|\textrm{no collisions in (0,t)})$, which could be called the conditional instantaneous collision probability (CICP).
However, the conditional distribution of $R(t)$, given no collisions in $(0,t)$, is not Gaussian, suggesting Monte Carlo or another approach is needed to compute the CICP.
Note also that this issue arises for collision probabilities but not for close approach probabilities since multiple close approaches can occur in an interval of time.

An advantage of computing the instantaneous collision probability with CFI is that it does not require analytical approximations.
The instantaneous collision probability as defined in \eqref{ICP} can be computed using CFI with high-accuracy.
However, the Gaussian uncertainty assumption is an approximation and therefore a source of error.
Characterizing the object state uncertainty with a Gaussian mixture model may reduce this error, but not eliminate it completely.
Monte Carlo is likely a more accurate, but more expensive, method for estimating instantaneous collision probabilities if Gaussian assumptions are not made.
The spherical geometry approximation is also a source of error when computing collision probabilities \citep{chan2018}.
This error is less concerning if a close approach, and not strictly a collision, is the event of interest.

\section{Conclusion}
\label{sec:conclusion}

This paper showed how to compute the instantaneous collision probability using a method called characteristic function inversion.
The approach is motivated by the observation that the squared separation distance has the generalized chi-square distribution under the assumption of Gaussian positional uncertainties.
Computationally, the approach requires an eigendecomposition of the combined covariance matrix and a numerical integration step, which is implemented in open-source \verb+Python+ and \verb+R+ packages.
An advantage of the procedure is that it does not require analytical approximations of the collision probability.
Furthermore, the approach has been established in the literature for over fifty years and can be used without modification for computing collision probabilities in the short-term encounter case.
A possible follow-on analysis is to extend the approach to evaluate the joint distribution of multiple squared separation distances.

\appendix

\section*{Appendix: Gaussian Mixture Model Generalization}
\label{app:gmm}

This appendix gives the instantaneous collision probability when the object positions are described by Gaussian mixture models.
The positions of the objects, conditioned on mixture components, are assumed to be jointly Gaussian distributed and independent.
For notation, suppose the first object has a Gaussian distribution with mean $\mu_{1,j}$ and covariance matrix $\Sigma_{1,j}$ with probability $w_{1,j}$ for $j=1,...,n_1$.
Similarly, suppose the second object has a Gaussian distribution with mean $\mu_{2,j'}$ and covariance matrix $\Sigma_{2,j'}$ with probability $w_{2,j'}$ for $j'=1,...,n_2$.
Then the relative position has a Gaussian distribution with mean
\begin{equation}
\mu_{j,j'}=\mu_{1,j}-\mu_{2,j'}
\end{equation}
and covariance matrix
\begin{equation}
\Sigma_{j,j'}=\Sigma_{1,j}+\Sigma_{2,j'}
\end{equation}
with probability $w_{1,j}w_{2,j'}$.
The weights, $w_{1,j}$ and $w_{2,j'}$, are positive and sum to one over $j$ and $j'$.

The instantaneous collision probability for this Gaussian mixture model is
\begin{align}
\label{icpgmm}
p(R'R<q^2)&=\sum_{j=1}^{n_1}\sum_{j'=1}^{n_2}w_{1,j}w_{2,j'}f(q,\mu_{j,j'},\Sigma_{j,j'}),
\end{align}
where $f(q,\mu,\Sigma)$ is the instantaneous collision probability for a distance $q$, mean $\mu$, and covariance matrix $\Sigma$.
Note that \eqref{icpgmm} is a weighted sum of collision probabilities for different combinations of components from each GMM.
The collision probability in \eqref{ICP} is obtained as a special case of this equation with $n_1=n_2=1$ and $w_{1,1}=w_{2,1}=1$.

\section*{Appendix: A Note on Notation}
\label{app:notation}

The notation in this paper differs from the typical notation used to characterize satellite collision probabilities.
Typically, the radius of the spherical conjunction domain is denoted $R$ and the bivariate Gaussian distribution in the STES is parameterized with mean $(x_m,y_m)$ and variances $(\sigma_x^2,\sigma_y^2)$.
Here, the radius of the spherical conjunction domain is denoted $q$ instead of $R$.
This choice lets $R$ denote the relative position, which is convenient since capitol letters are often used to represent random variables, such as the relative position in this case.
For the Gaussian distribution parameters in the STES, the mapping between notations is $\lambda_1=\sigma_x^2$, $\lambda_2=\sigma_y^2$, $\nu_1=x_m/\sigma_x$, and $\nu_2=y_m/\sigma_y$.
The $(\lambda,\delta)$ parameterization is used here to align with software for evaluating the generalized chi-square distribution.

\newpage

\section*{Disclaimer}

\begin{quote}
This document was prepared as an account of work sponsored by an agency  of the United States government. Neither the United States government nor Lawrence Livermore National Security, LLC, nor any of their employees makes any warranty, expressed or implied, or assumes any legal liability or responsibility for the accuracy, completeness, or usefulness of any information, apparatus, product, or process disclosed, or represents that its use would not infringe privately owned rights. Reference herein to any specific commercial product, process, or service by trade name, trademark, manufacturer, or otherwise does not necessarily constitute or imply its endorsement, recommendation, or favoring by the United States government or Lawrence Livermore National Security, LLC. The views and opinions of authors expressed herein do not necessarily state or reflect those of the United States government or Lawrence Livermore National Security, LLC, and shall not be used for advertising or product endorsement purposes.
\end{quote}

\begin{quote}
Lawrence Livermore National Laboratory is operated by Lawrence Livermore  National Security, LLC, for the U.S. Department of Energy, National Nuclear Security Administration under Contract DE-AC52-07NA27344.
\end{quote}

\bibliographystyle{apalike}
\bibliography{ms}

\end{document}